# Post trimming of silicon photonics microresonators by nanoscale flash memory technology

Meir Grajower[1], Noa Mazurski[1], Joseph Shappir[1] and Uriel Levy[1]

[1]Department of Applied Physics, The Benin School of Engineering and Computer Science, The Hebrew University of Jerusalem, Jerusalem, 91904, Israel

**Abstract**

Flash memory technology is widely common in modern microelectronics, and is essentially affecting our daily life. Considering the recent progress in photonic circuitry, and in particular silicon photonics circuitry, there is now an opportunity to embed the flash memory technology in photonic applications. A particularly promising candidate that can benefit from such integration is the photonic resonator. As of today, chip scale resonators are essential building blocks in modern silicon photonic platform. However, their properties, and in particular their resonance frequencies deviate from their designed values due to unavoidable fabrication imperfections, imposing a stringent limitation on the applicability of such devices. Here we present a solution for this major obstacle and demonstrate electrical approach for post trimming of such resonators. This is achieved by integrating the well-established flash memory technology into the photonic circuitry. More specifically, we use the nanoscale Silicon-Oxide-Nitride-Oxide-Silicon (SONOS) structure in order to trap charges in the thin silicon nitride layer, which is located in close proximity to the silicon device layer. This enables the accumulation of charges in the silicon, modifying the effective index of the optical mode and consequently the resonance frequency. By doing so, we provide a robust and elegant CMOS compatible memory solution, which can be easily manufactured and commercialized. We expect such an approach to pave the way for even more efficient utilization of resonators and interferometers in chip scale photonic and electro optic systems, with wide range of applications such as filters, modulators, sensors, and lasers, to name a few.

**Introduction**

Silicon based nanophotonic devices such as interferometers and resonators are serving as important building blocks in modern chip scale photonic circuits, supporting diverse functionalities such as modulation, switching, filtering, wavelength selection and dispersion control, biochemical sensing and others[1–13]. Integrating silicon with advanced platforms such as conductive oxide, polymers and atomically thin two dimensional materials provide a route for even more advanced silicon based active nanophotonic devices [14–20].

A major challenge related to the practical implementation of nanophotonic resonators in chip scale optical communications systems is related to the accuracy in setting the resonance wavelength. For example, the actual resonance wavelength of two adjacent resonators on the same chip can easily fluctuates in the range of about 10GHz due to fabrication imperfections, (see e.g. [21]). Such fluctuations can be compensated by controlling the refractive index of the medium, either actively, e.g. by the use of the thermo optic effect [7,22–24], or passively, by the use of post trimming approaches. Clearly, the thermo optic approach is far from being ideal as heating the structure consumes significant electrical power. In addition, maintaining a constant difference in temperature between two adjacent resonators on a chip is challenging. Post processing trimming approaches which does not consume power are thus preferred. Indeed, several such approaches have been demonstrated, e.g. the use of polymers, locally oxidizing the silicon surface and phase change materials [25–28]. Yet, there is still a need for a robust, reliable, accurate, reproducible and large scale CMOS compatible approach for the post processing trimming of silicon nanophotonic structures.

Hereby, we propose and experimentally demonstrate a novel solution for the post processing trimming of nanophotonic resonators in silicon, by integrating the well-established SONOS (Silicon-Oxide-Nitride-Oxide-Silicon) flash memory technology into the photonic chip. The approach is based on the trapping of charges in a thin nanoscale layer of silicon nitride embedded between two silicon oxide layers. The device is operating as a MOS (Metal-Oxide-Semiconductor) capacitor, and the trapped charge modifies the free carriers concentration in the silicon device layer of an SOI (silicon on insulator) based nanoscale waveguides, which are used as the basic building block of chip scale resonators. The end result is an **electrically controlled** memory which shifts of the resonance frequency via the free carrier plasma dispersion effect.

**Fabrication**

The device was fabricated on a commercial SOI wafer (SOITEC) with a 2 μm buried oxide layer and a 220nm top silicon (Boron doped, $N_a = 5\text{x}10^{17}[\text{cm}^{-3}]$) layer. The silicon waveguides were implemented in a rib configuration with cross sectional dimensions of 450 nm width, 220 nm height and rib of 80nm (Fig. 1a). Two microring resonators (MRRs, 30μm diameter each) were coupled to the bus waveguide in series, one after the other. This structure was fabricated in a commercial fab (Towerjazz semiconductor Ltd.). On top of the waveguide and MRR structures, three insulating layers were formed: 8nm of silicon oxide by thermal oxidation, 6nm of plasma enhanced chemical vapor deposition (PECVD) silicon nitride a ~500nm thick PECVD silicon oxide. On top of the thick oxide above the MRR, a 150nm thick layer of aluminum (Al) was deposited using electron beam evaporation to form the gate of a MOS structure as shown in Fig 1. The Al layout above the two MRRs is not identical - above MRR C1 (Fig 1d), the Al contact pad area is small (~10,000μm$^2$) and only covers the silicon MRR partially. This aluminum is patterned as a periodic structure with a period of 2μm and duty cycle of 0.5. These openings enable photo emission of electrons from the silicon using ultra-violet (UV) light source, as will be explained later. Above MRR C2 (Fig 1d), the Al covers the entire MRR with a larger contact pad area (~50,000μm$^2$). This large pad serves two purposes: a) preventing photo emission by blocking the UV illumination, and b) eliminating the need to form Ohmic contact to the silicon layer. Additional contact is made in the back side of the SOI wafer. The obtained structure is that of a Metal-Oxide-

Semiconductor (MOS) capacitor with a silicon nitride layer sandwiched between the two silicon oxide layers near the silicon, forming an oxide-nitride-oxide (ONO) structure.

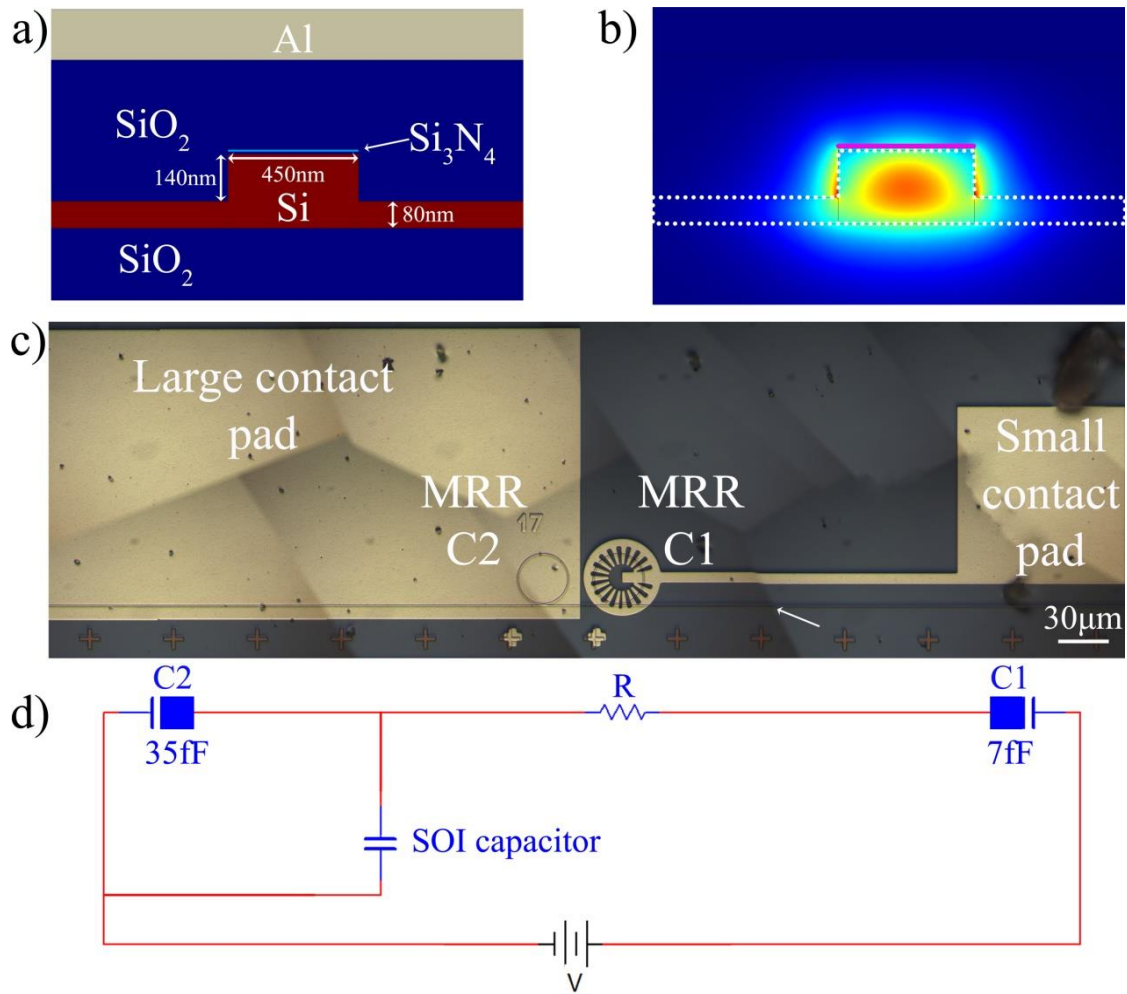

**Figure 1** - a) schematic cross section of the silicon waveguide. The Al contact on the back side of the SOI is not illustrated. b) cross section showing the electric field intensity of the transverse electric (TE) like mode supported by the waveguide. The white dotted line bounds the silicon region. The pink line represents the nitride layer. c) Panoramic microscope image of the waveguide and the two MRRs. The bright color corresponds to the Al. The waveguide and the MRRs can be clearly observed. d) Electrical scheme of the two MRRs and Al contacts. The silicon rib between the two MRRs is modeled as a resistor. The SOI capacitor is huge compared to the two MRR capacitors. The MRR C1 is only partially covered by Al to enable exposure to UV light. The MRR C2 is fully covered by a large Al electrode to prevent any light exposure. Furthermore, the circuit configuration 1d, allows the application of external voltage to MRR C1 without needing an Ohmic contact to the silicon layer.

**Concept of operation**

As mentioned before, the motivation of this work is to introduce an electrical post processing method of controlling the resonance frequency of the MRR. This goal is achieved by using the concept of the SONOS flash memory. In short, SONOS allows the injection of controlled amount of electrons into the nitride layer. These charges induce opposite charges in the silicon near the silicon-oxide interface and allow to drive the MOS capacitor into accumulation, depletion, and even inversion mode.

While in flash memory applications the readout of memory is achieve by measuring electrical parameters such as voltage or current, here we use optical readout instead. The induced change in the amount of free carriers results in a change in the effective refractive index of the MRR via the free carrier plasma dispersion effect[29]. As a result, the resonance frequency varies as well following the relation:

$$2\pi r n_{eff} = m\lambda \qquad (1)$$

At this point, it is helpful to review the basic physical concepts underlying the operation of the SONOS flash memory. Upon the application of positive voltage to the gate electrode of the device, electrons tunnel from the silicon through the oxide into the nitride layer. The relatively thin layers used in commercial SONOS based devices enable the electron tunneling at relatively low voltages. The gate electrode, which is about ~20nm away from the silicon, can be used both for the tunneling process as well as for the measurement of the amount of trapped charges.

For optical applications, the above mentioned thickness of the MOS insulating layer of about 20nm the is not sufficient due to the Ohmic loss in the metal. To avoid excessive optical loss, the thickness of the top oxide layer above the nitride layer was increased to ~500nm. This allows to minimize the effect of the Al gate layer on the loss of the waveguide mode, and thus avoiding a drastic reduction in the quality factor of the MRR. On the downside, the direct impact of increasing the thickness of the oxide layer is the need for operating at higher voltage during the charge trapping\detrapping. To avoid breakdown of the oxide layer, we replaced the tunneling process by internal photo emission as the method of injecting electrons into the nitride layer. The process is shown schematically in Fig. 2.

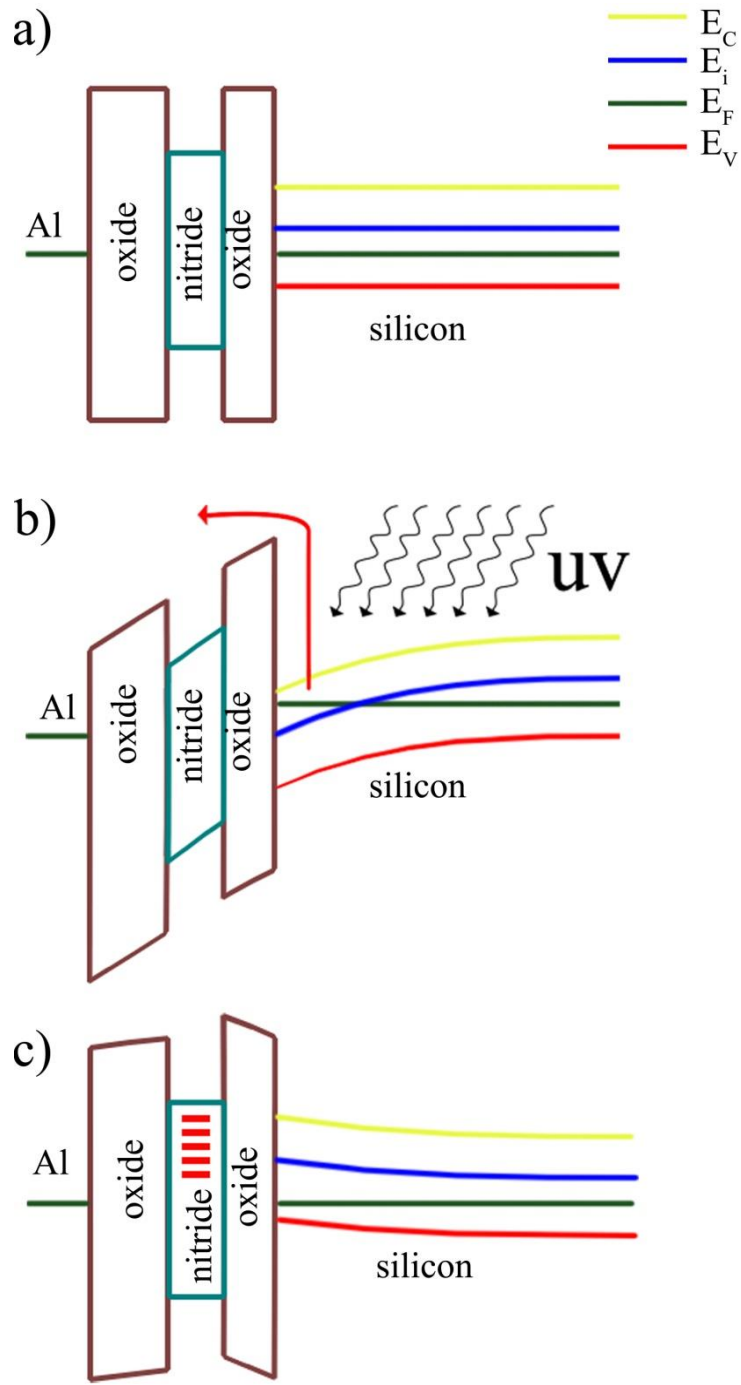

**Figure 2** – Schematic illustration of the charge trapping mechanism in our device (not to scale) a) flat band, i.e. no band bending. b) electron photo emission – achieved by a combination of UV illumination and the application of positive voltage. The red arrow denotes the photoemission of electrons from the valance band of the silicon into the silicon nitride. c) The band diagram after electrons are trapped in the nitride and without the application of external voltage. The electrons in the nitride (shown as red bars) attract holes from the silicon- oxide interface, and the capacitor is found in accumulation mode.

**Charge trapping**

Unlike tunneling injection from the silicon conduction band (as in the Flash memory case), in our case of internal photo emission the electrons are injected from the silicon valence band. This is due to the low probability of the photo emission process of electrons from the conduction band. The potential barrier between the silicon valence band and the silicon oxide conduction band is 4.3eV. This high energy was supplied by a UV light emitting diode with center wavelength around 250nm (4.96eV).

To ensure that the UV illumination penetrates all the way to the silicon, we replaced the continuous metal electrode above the MRR by a comb like structure. The amount of trapped electrons is determined by the applied positive gate voltage. The photo exited electrons at the silicon surface layer are carried towards the nitride layer by the electric field of the applied gate voltage. This charge injection is a self-stopping process as the trapped electrons induce an opposite electric field with increasing strength which is directly proportional to the trapped charge density. This increasing field reaches a value which compensates the gate induced electric field to the point that photo induced electron injection is stopped. Thus, the saturation level of the injected electrons is determined by the value of applied gate voltage.

**Optical effect of the trapping**

Once the desire amount of trapped electrons is reached, the gate voltage is removed. The trapped charges are now neutralized by opposite charges divided between the two electrodes of the MOS capacitor. Yet, considering the fact that the silicon electrode is located only 8 nm away of the nitride, whereas the metal gate is separated by a large distance of ~500nm, in practice all the compensating positive charge is located in the silicon.

The compensating charge is induced either by generating an accumulation layer at the interface (for p-type silicon) or by depleting free carriers (in the case of n-type silicon). The change in free carrier concentration induces local change in the refractive index of silicon according to[29]:

$$\Delta n = -(8.8x10^{-4}\Delta N_n + 8.5\Delta N_h^{0.8}) \cdot 10^{-18} \qquad (2)$$

Where $\Delta N_n$ is the change in electron concentration, and $\Delta N_h$ is the change in hole concentration. Here, we have used a p-type silicon waveguide. As a result, an accumulation layer is generated at the interface between the silicon and the oxide layer, inducing a local change in the refractive index which in turn controls the resonance frequency of the MRR.

As shown in Fig 1, the device is constructed of two MRRs connected electrically in series. MRR C1 (Fig 1c,d) has a much smaller capacitance relative to MRR C2. Thus, the applied voltage on the two MOS capacitors in series falls mostly on C1, and the trapped charges are located on top of MRR C1. The end result is that MRR C2 serves as reference which allows to directly and accurately measure the resonance shift of MRR C1 by reducing the influence of thermal effects. In addition, this configuration eliminates the need for Ohmic contact to the silicon. Two MRR configuration was recently used for the purpose of ultraprecise sensing [21].

**Results**

**Experimental setup**

The measurement setup consists of a tunable laser source (Agilent 81640A) that is coupled in and out of the device using lens fibers in a butt coupling configuration. The transmission of the device is measured as a function of wavelength. In addition, a variable DC power supply (SRS PS310) is connected to two probes which applied voltage on the MRRs (connected to gate of MRR C1 and MRR C2, Fig 1c and 1d). The gate electrode of MRR C2 (large area) together with the back side of the SOI are connected to the ground potential. The voltage is applied to the gate metal of MRR C1. As mentioned, based on the difference in the capacitance of C1 and C2, practically almost all the voltage is applied to MRR C1.

Electrons are photo injected from the silicon into the nitride layer by UV illumination (UV LED, Thorlabs E250) where positive DC voltage is applied on the gate of MRR C1.

**Experimental results**

The two MRRs are fabricated with the same dimensions, and are thus expected to have the same optical behavior. However, it is well known that local minor variations in the fabrication process parameters lead to variations in the resonance frequency[21]. Indeed, the measured transmission spectrum (Fig 3) shows clear traces of two resonances originating from the two MRRs which are slightly detuned.

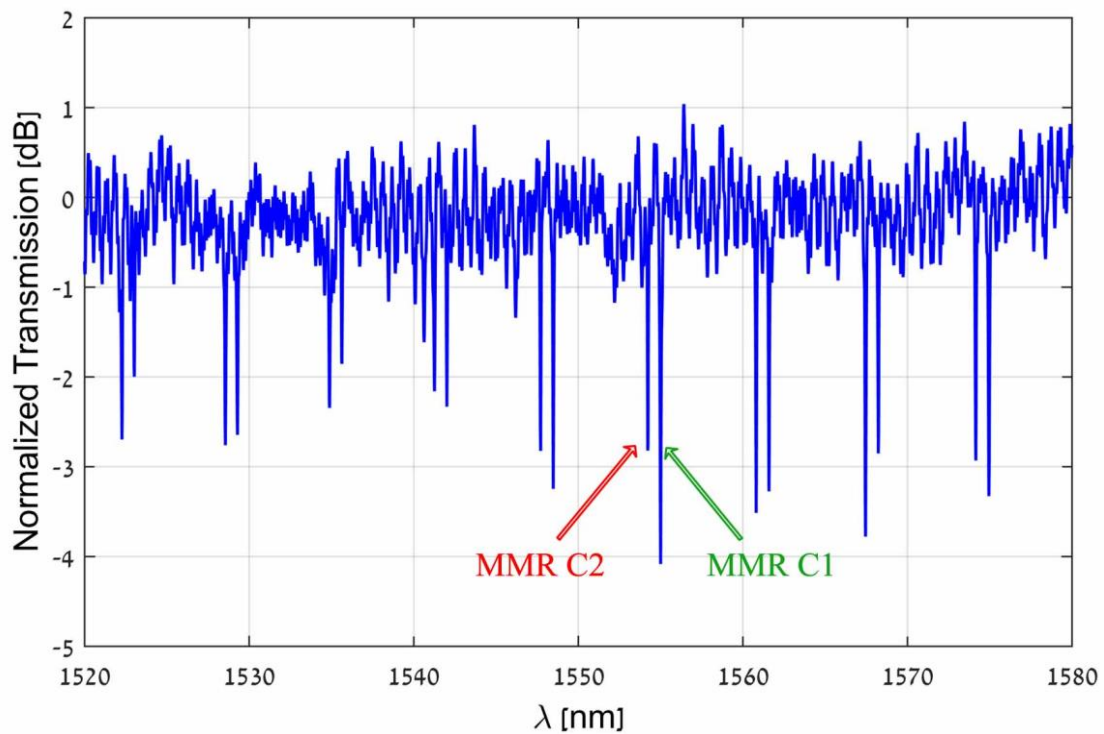

**Figure 3** - Normalized transmission spectrum of the device. Even though the parameters of the two MRRs are identical, two different resonance frequencies corresponding to the two MRR are evident.

Next, the devices were tested for their response to variable gate voltages, without the application of UV light, so that charge trapping is still not active. The effect of positive and negative gate voltages is shown in Fig. 4 where we limit the scan to two adjacent resonances.

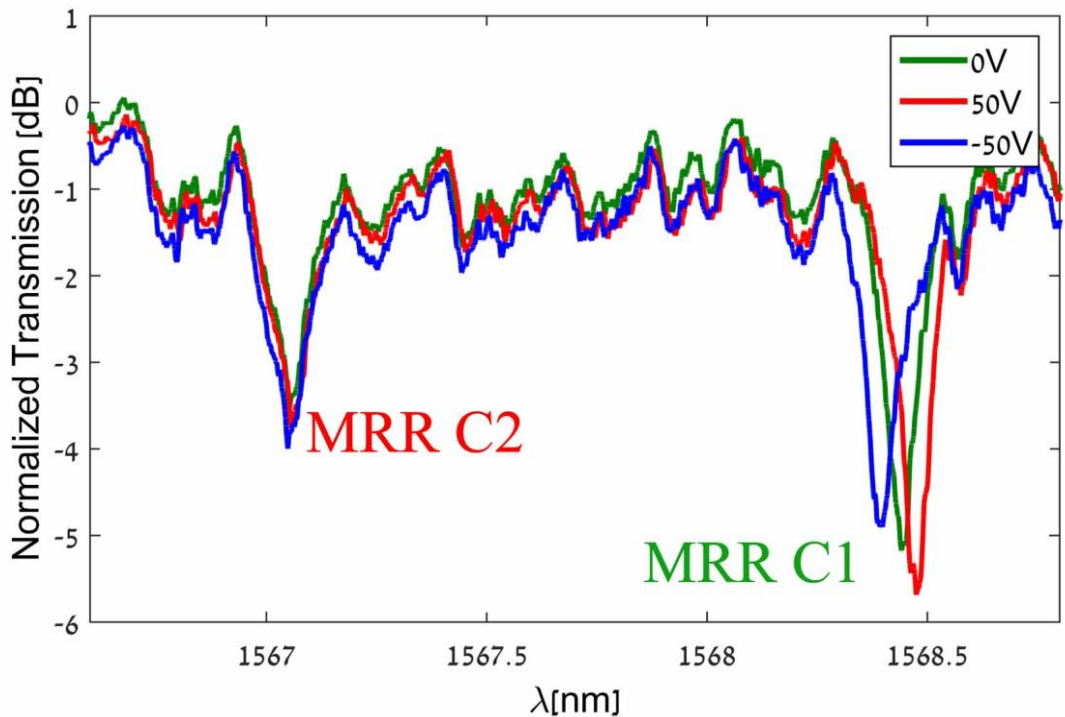

**Figure 4** – Normalized transmission spectrum of the device measured under three different applied voltages. As expected, under positive voltage the spectrum is red shifted while under negative applied voltage it is a blue shifted.

As observed from Fig. 4 (right resonance, MRR C1), blue shift is obtained for negative voltage while red shift is obtained for positive voltage. This is expected, because for negative voltage holes are accumulated, reducing the effective refractive index, whereas for positive voltage the holes are depleted and the effective refractive index is increased. It is also clear from Fig. 4 that the left side resonance shows no shift under applied bias, and is therefore attributed to MRR C2, while the right resonance corresponds to MRR C1 (this is designated by the arrows in Fig. 3).

Typically, a MOS capacitor is not in a flat band condition under zero applied bias. This is because of the initial difference in work function of the metal and the silicon, as well as due to trapped charges in the dielectric layer during the fabrication process. As a result, an initial band bending is expected. In order to characterize the band bending, we repeated the experiment presented in Fig. 4, by performing spectral measurement under different positive voltage bias, with smaller voltage steps and extracted the resonance shift at each applied voltage. The obtained results are shown in Fig. 5 (red curve). In additions, the expected resonance shift was calculated (blue curve) using FEM solver (COMSOL Ldt., see supplementary). To fit calculations to the experimental results, the following procedure was adopted. First, the resonance shift was divided by 1.8. This can be explained by the fact that the model assumes uniform voltage across the MRR whereas in practice the metal gate is patterned (in a fingers shape) and covers only 50% of the MRR area. Given the fact that the fingers width and spacing of the gate metal are twice the thickness of the gate oxide so that the effect of fringing fields cannot be ignored.

Next, to fit the voltage axis of the simulation to the measured data, two parameters were used - shift and scaling. The shift is attribute to the initially trapped charges in the oxide and nitride (as mentions before), while the scaling is attribute to variation of the oxide thickness and dielectric constant[30].

From the shape of the curves and from extracting these fitting parameters, it can be understood that at zero gate voltage there is already an accumulation layer at the silicon surface. As the voltage is increased, the accumulation layer is diminished and flat band condition is achieved. Further increase of the voltage induces the formation of depletion layer which reaches its maximum at gate voltage of about 80V. As the gate voltage increases beyond ~80V, the resonance frequency shift is decreased, indicating that an inversion layer is formed. Therefore, the flatband voltage can be calculates by: $V_{FB} = V_G - V_T = 80 - V_T = 80 - \left(2\phi_F + \frac{Q_d}{C_{ox}}\right)$, where $V_{FB}$ is the flat band voltage, $\phi_F$ is the Fermi energy, $Q_d$ is the depletion charge, $C_{ox}$ is the thick oxide capacitance and $V_T$ is the threshold voltage for inversion. Using this relation, the flatband voltage is found to be ~45V. The relatively high values of $V_T$ as compared to standard MOS devices is attributed to the thick gate oxide as well as to the trapped charges in the nitride layer as it comes out of the fabrication process.

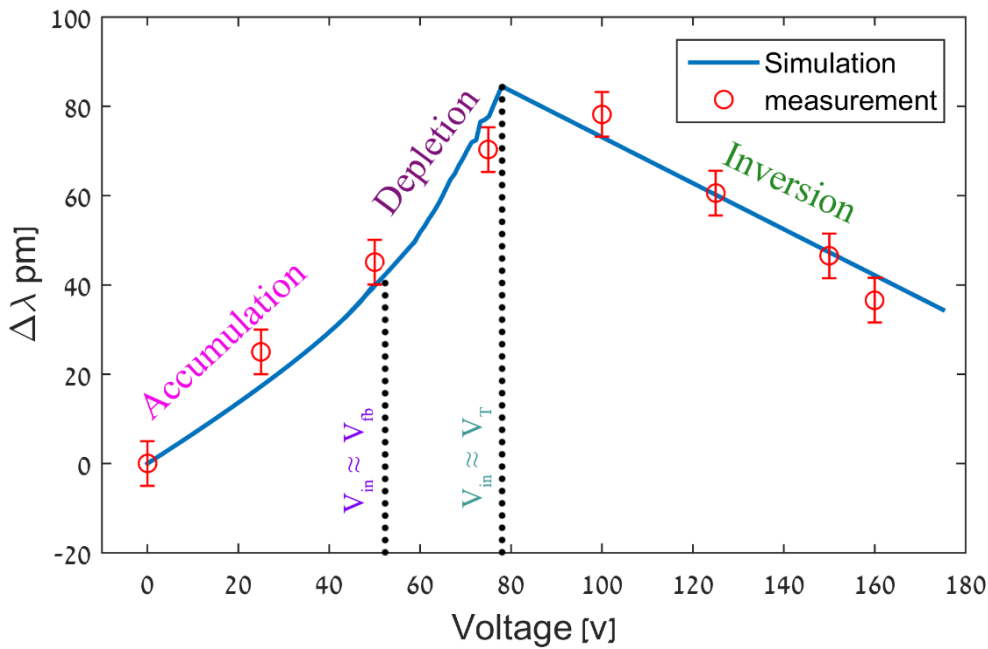

**Figure 5** - Simulation results (blue line) and experimental measurements (red circles) showing the shift between the resonance frequencies of the two MRRs under different applied voltage. The resonance frequency shift at 0V is set to 0. Also indicated are the different regimes of the silicon band bending.

Following the characterization of our device as a MOS capacitor, we turn into demonstrating the major goal of this paper, namely the possibility of obtaining permanent and controlled shift of the resonance wavelength by charge trapping in the thin nitride layer embedded within the MOS dielectric as is commonly done in Flash memories. As previously mentioned, we increased the total dielectric thickness from ~30nm to nearly 500nm in order to reduce the propagation loss of the optical mode which results in from the Ohmic loss in the metal. Such a thick oxide, which is deposited using PECVD, is exposed to defects and cannot sustain high electric fields. As a result, the overall voltage that can be applied is limited by the breakdown of the thick oxide, effectively limiting the electric field that can be applied to the thin tunneling oxide layer separating the silicon from the silicon nitride layer. This prevented us from using tunneling effect to transfer electrons from the silicon into the nitride layer. To overcome this obstacle, we choose the use of internal photo emission, i.e. UV assisted transfer of electrons from the valance band of the silicon to the silicon nitride layer.

Similar to the influence of the gate voltage on the resonance position (Figs 4, 5), Fig. 6a demonstrates the effect of the trapped charge in the nitride on the resonance frequency. Unlike the previous case, here the device was measured at zero gate voltage, after injecting charges at three different gate voltages. As can be seen, a noticeable trapping effect is observed. To better quantify this effect, we have repeated the charge trapping at many gate voltages. After each trapping procedure we measured the transmission spectrum of the device at zero gate voltage and extracted the resonance shift for each of the trapping events. The quantitative dependence of the resonance shifts on the trapped charges which are expressed through the gate voltage during the photo emission step is shown in Fig. 6b (red circles). We have applied an increasing voltage up to 170V. When we further increased the voltage, an electrical breakdown occurred (See supplementary S2). Finally, we calculated the shift of each trapping event (blue curve), using the same fitting parameters as in Fig. 5.

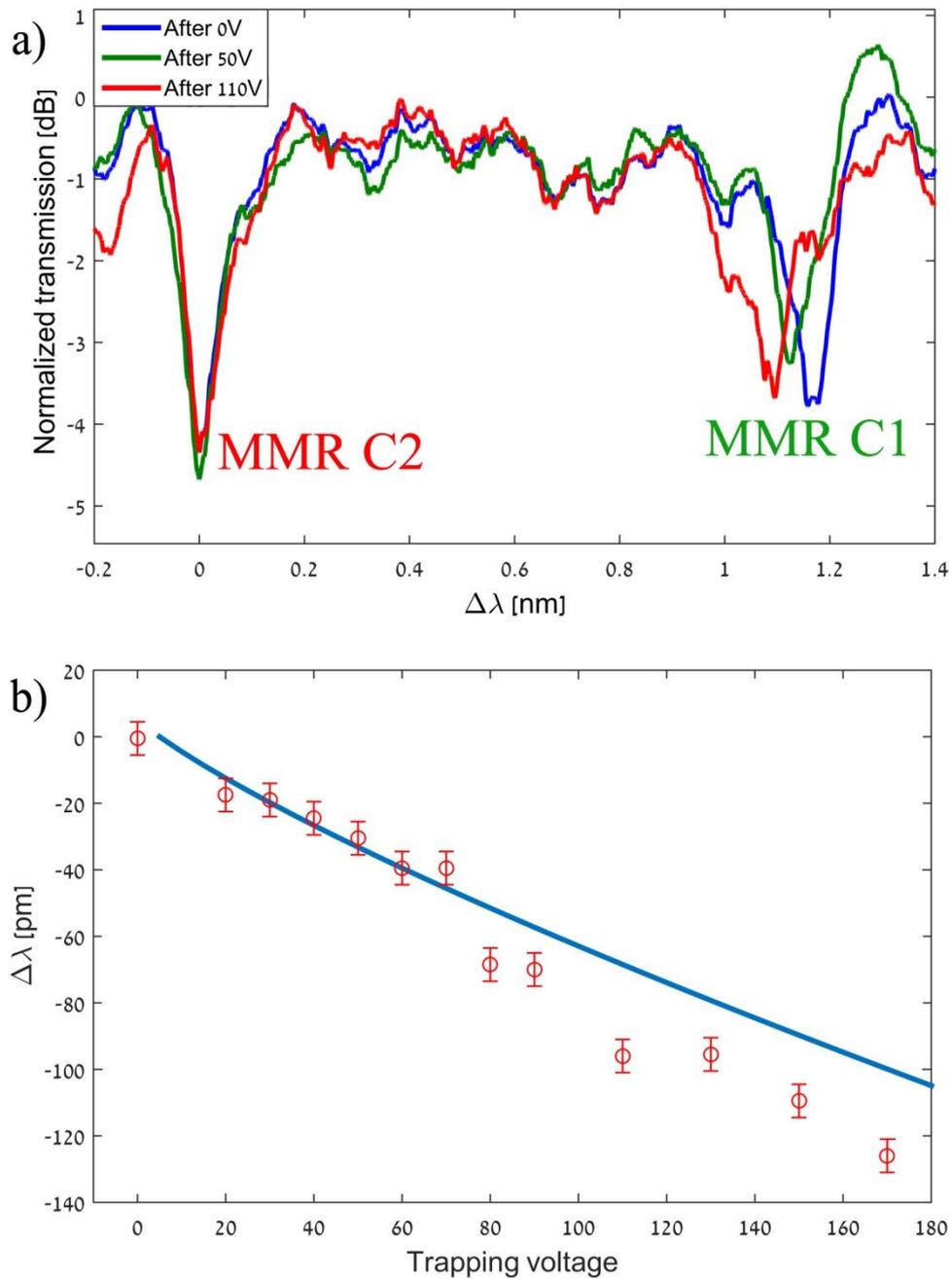

**Figure 6** – Experimental measurement demonstrating the effect of the trapped charge in the nitride layer on the resonance frequency of the MRR. a) Transmission spectrum measured at 0V after UV illumination at three different gate voltages b) Resonance shift measured (red circles) and calculated (blue line) at 0V after UV illumination under increasing applied gate voltages.

As previously mentioned, the shift of the resonance frequency is dependent of the applied gate voltage as follow. The maximal charge that can be trapped in the nitride is estimated by $Q_{max} = C_{ox}V_G$ (where $V_G$ is the gate voltage and $C_{ox}$ is the thick oxide capacitance). After the trapping event, the gate voltage is remove and the trapped charges induce an effective gate voltage in the silicon. The effective voltage is found to be $V_{eff} = Q_{max}/C_{ono}$ where $C_{ono}$ the capacitance of the thin oxide layer separating the nitride from the silicon. In our case, the effective voltage can be found to be $V_{eff} = V_G \frac{C_{ox}}{C_{ono}} \cong \frac{V_G}{65}$. Thus, the maximal effective field is the range of ~0.3 volt per nanometer, which is the same order of magnitude as demonstrated in accumulation-based silicon photonic modulators [31]. For such a field strength in the trapping process we obtained a resonance shift of about 125pm. This value can be further increased, e.g. by using a polysilicon electrode separated from the nitride by a high quality oxide layer which will allow both applying stronger fields, up to the breakdown field (~1 volt per nm), and also reducing the operation voltage for the trapping procedure. Such an approach may also allow to work in tunneling mode. This will enable not only trapping, but also the opposite action, i.e. erasing the trapping by applying an opposite voltage.

So far, we have shown separately the effect of gate voltage and trapped charge on the resonance shift. Next, we present the cumulative effect of the trapped charge and the applied gate voltage. In Fig. 7 we show the resonance shift for different gate voltages after significant charge trapping at 160V (red circles). For comparison, we also show the resonance shift of the same device at the same gate voltage prior to the charge trapping process (blue circles) before charge trapping (as in Fig. 5). For simplicity, the initial frequency shift between the two MRRs is defined as zero in both cases, even though the shift is obviously different due to the trapping process (see Fig. 6b).

As can be observed, in the case where no intentional charge trapping is applied, the threshold voltage, $V_T$, is achieved when the charge in the MOS capacitor compensates the initial charges in the nitride (which is typically fabrication process dependent) and in the depletion layer. A voltage of ~80V is needed to achieve this condition. However, in the case where we intentionally trapped charges (red circles), the additional negative trapped charge injected into the nitride at $V_G=160V$ result in a significant increase in the total negative charge trapped in the nitride up to the point where even a gate voltage as high as 160V is not sufficient to compensate the negative charge in the nitride and in the depletion region. Consequently, $V_T$ is not reached, and a major difference in operation conditions between the pre-trapped and the post trapped cases is obtained.

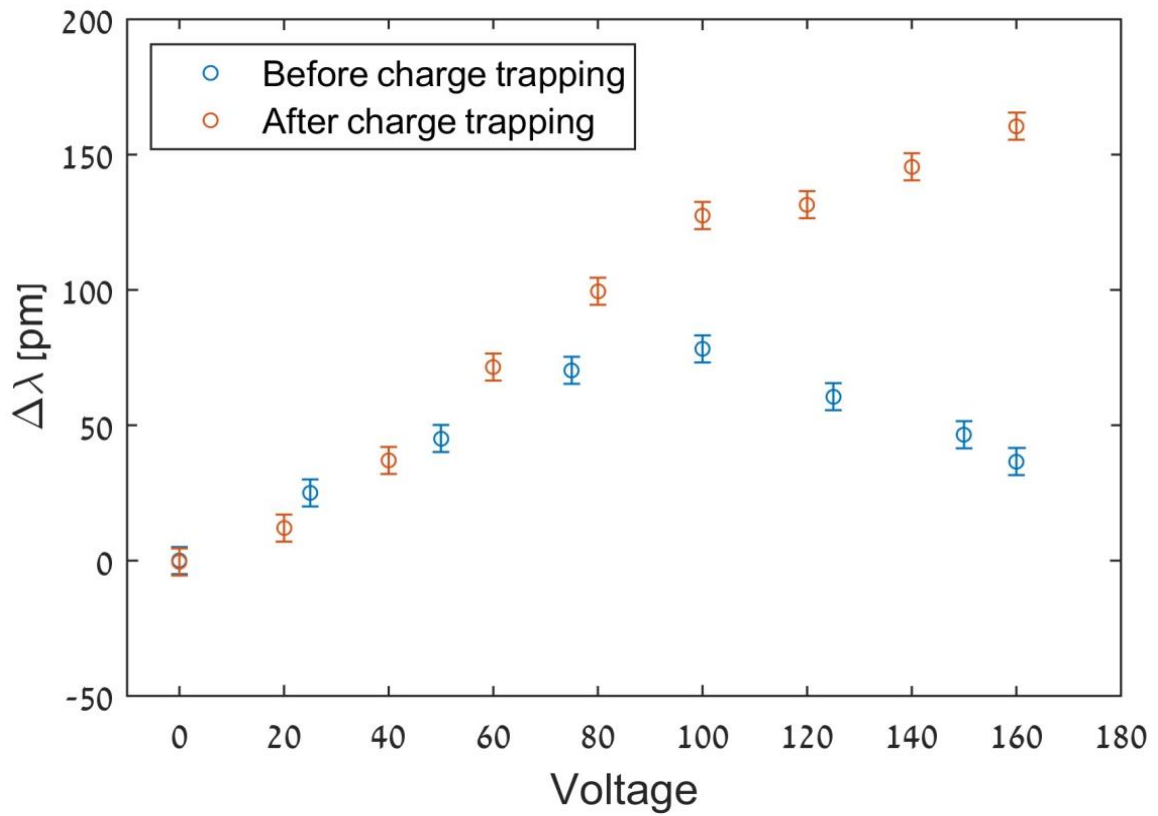

**Figure 7** - measurement of the resonance shift due to applied gate voltage after charge trapping (red circles). Also shown are the results before charge trapping, as also shown in Fig. 4 (blue circles). For simplicity, the resonance shift between the two MRRs at zero gate voltage is set to zero in both cases, even though the shift is different due to the trapping operation.

**Discussion and Conclusions**

In conclusion, we have integrated a SONOS based flash memory technology into photonic circuitry and utilized this structure to demonstrate a novel method of post process resonance trimming of microresonators in silicon photonics platform. A wide range of post processing trimming values, up to about 125 pm in resonance wavelength were demonstrated. This approach is based on controlled injection of electrons into a thin nitride layer bounded between two oxide layers in the insulating layer of MOS structure. As of today, the flash memory is a very common used and it is being manufactured in high volumes at low cost. Thus, the proposed approach allows easy fabrication in a commercial CMOS fab and offers high reliability over time. Indeed, we have observed a stable charge trapping with no noticeable variations in resonance shift over the course of few weeks, and based on its proven reliability it is expected that undesired degradation in trimming capability will not be observed even at longer duration of time (years).

The demonstrated concept can readily be used for overcoming unavoidable fabrication inaccuracies in a silicon photonics fab and to align the resonance frequency of chip scale photonic resonators on demand. While we have demonstrated the use of this method in microring resonators (MRR), this approach of using flash memory for achieving a controllable and permanent change of the refractive index of silicon, can be implemented also in other photonic configurations such as photonic crystals, Mach Zehnder interferometers, Bragg gratings and other photonic devices operating on the basis of resonance or interference effect.

The demonstrated maximum shift in resonance wavelength of about $\Delta\lambda_{max} = 125[pm]$ is already a significant landmark as it is typical of the fluctuations in resonance frequency between various resonators on the same chip, driven by slight variations in manufacturing.

Yet, it should be possible to increase the tunability even further, e.g. by better aligning the gate electrode to the resonator and by using more advanced electrode design. Consequently, much larger fraction of the mode area in the MRR is expected to become exposed to the UV radiation, resulting in larger shift of the resonance frequency. Furthermore, one may replace the metal gate electrode with a poly-crystalline silicon layer. This will enable locating the gate electrode much closer to the silicon as is the case in flash memories (~40nm) and will allow charge to be directly injected by the application of low voltage, without the need for UV illumination. Furthermore, it will allow the erasing of charges as well, paving the way for extending the functionality of the device. For example, devices such as CMOS compatible electrically write – optically read memories could be implemented, offering a competitive alternative to previously reported non CMOS compatible approaches such as plasmonic memristors and others[32–34]. Given the wealth of potential applications, combined with the ease of fabrication and the high potential for commercialization makes the integration of the SONOS flash memory technology into silicon photonics circuitry makes it a particularly promising approach for the future development in fields related to silicon photonics integration.

**Acknowledgements**

The authors thank Dr. Boris Desiatov for fruitful discussions, and Nachi Vofsi for the fabrication of the micro resonators. This work was supported by the Kamin program of the Israeli ministry of industry and trade. Meir Grajower acknowledges a Ph.D fellowship from the center for nanoscience and nanotechnology of the Hebrew University.